\documentclass[fleqn,10pt]{wlscirep}
\usepackage[super, comma, sort&compress]{natbib}
\usepackage[utf8]{inputenc}
\usepackage[T1]{fontenc}
\usepackage{bm}
\usepackage[tableposition=top]{caption}
\usepackage{tcolorbox}
\usepackage{multirow}
\usepackage{changepage,threeparttable}
\usepackage{makecell}
\usepackage{lipsum}
\usepackage{booktabs}
\usepackage{enumerate}
\usepackage{svg}
\setcitestyle{super,comma,open={[},close={]}} 
\title{Structure-based drug discovery with deep learning}

\author[1,2+]{R{\i}za \"{O}z\c{c}elik}
\author[1,2+]{Derek van Tilborg}
\author[3]{José Jiménez-Luna}
\author[1,2*]{Francesca Grisoni}
\affil[1]{Eindhoven University of Technology, Institute for Complex Molecular Systems and Dept. Biomedical Engineering, Eindhoven, Netherlands.}
\affil[2]{Centre for Living Technologies, Alliance TU/e, WUR, UU, UMC Utrecht, Netherlands.}
\affil[3]{Microsoft Research Cambridge, Cambridge, United Kingdom.}

\affil[+]{These authors contributed equally.}

\affil[*]{e-mail: \href{mailto:f.grisoni@tue.nl}{f.grisoni@tue.nl}}

\begin{abstract}
Artificial intelligence (AI) in the form of deep learning bears promise for drug discovery and chemical biology, \textit{e.g.}, to predict protein structure and molecular bioactivity, plan organic synthesis, and design molecules \textit{de novo}. While most of the deep learning efforts in drug discovery have focused on ligand-based approaches, structure-based drug discovery has the potential to tackle unsolved challenges, such as affinity prediction for unexplored protein targets, binding-mechanism elucidation, and the rationalization of related chemical kinetic properties. Advances in deep learning methodologies and the availability of accurate predictions for protein tertiary structure advocate for a \textit{renaissance} in structure-based approaches for drug discovery guided by AI. This review summarizes the most prominent algorithmic concepts in structure-based deep learning for drug discovery, and forecasts opportunities, applications, and challenges ahead.
\end{abstract}

\begin{document}
\newcommand{\Equref}[1]{Equation~(\ref{#1})}
\newcommand{\equref}[1]{Equation~(\ref{#1})}
\newcommand{\Secref}[1]{Section~\ref{#1}}
\newcommand{\secref}[1]{Section~\ref{#1}}
\newcommand{\Figref}[1]{Figure~\ref{#1}}
\newcommand{\figref}[1]{Figure~\ref{#1}}
\newcommand{\dashfigref}[2]{Figures~\ref{#1}--\ref{#2}}
\newcommand{\Tabref}[1]{Table~\ref{#1}}
\newcommand{\tabref}[1]{Table~\ref{#1}}
\newcommand{\TODO}[1]{\textbf{\textcolor{red}{#1}}}
\newcommand{\todo}[1]{\textbf{\textcolor{red}{#1}}}
\newcommand{\note}[1]{\textbf{\textcolor{blue}{#1}}}
\newcommand{\denovo}{\textit{de novo}}
\newcommand{\etal}{\textit{et al.}}
\newcommand{\riza}[1]{\note{Rıza: #1}}

\flushbottom
\maketitle

\subsection*{Keywords}
Artificial intelligence $\bullet$ De novo design $\bullet$ Machine learning $\bullet$ Medicinal chemistry $\bullet$ Structural biology

\thispagestyle{empty}

\section*{Introduction}

Deep learning -- a subfield of artificial intelligence (AI) based on multi-layer neural networks\cite{lecun2015deep} -- has gained remarkable traction in science and technology, \textit{e.g.}, to advance mathematics \cite{davies2021advancing,fawzi2022matrix}, investigate galaxies \cite{ho2022dynamical}, and generate realistic images\cite{gregor2015draw}. Chemistry and biology have witnessed several AI breakthroughs, for instance, in protein structure prediction\cite{jumper2021highly,baek2021accurate}, chemical synthesis planning\cite{segler2018planning, coley2017prediction}, and atomistic simulations\cite{batatia2022mace, batzner2022}. Drug discovery has particularly benefited from the advent of deep learning\cite{jimenez2021artificial, chen2018rise}, with success in molecule prioritization and automated \textit{de novo} design\cite{stokes2020antibioticsgraphs, segler2018generatingfocused,yuan2018chemicalspacemimicry, merk2018denovo}. Here, deep learning can accelerate the navigation of the extremely vast chemical space of drug-like molecules\cite{dobson2004chemical} in search for potential therapeutics, and complement resource- and time-intensive high-throughput screening campaigns \cite{schneider2018automating}.
Most deep learning studies have focused on ligand-based approaches \cite{jimenez2021artificial}, which leverage solely the structural information of small molecule ligands to provide predictions. For these applications, numerous systematic studies \cite{wu2018moleculenet,brown2019guacamol} and experimental proofs-of-concept\cite{yuan2018chemicalspacemimicry,merk2018denovo,zhavoronkov2019deep} have been published. On the other hand, structure-based deep learning approaches -- which leverage information on the target protein -- have not found parallel interest yet.

Structure-based drug discovery (SBDD) methods augmented with AI are arguably a more complex and a higher-potential endeavor compared to their ligand-based counterparts. Numerous marketed drugs have been identified by `traditional' SBDD (\textit{e.g.}, HIV-1 protease inhibitors\cite{wlodawer1998inhibitors}, the thymidylate synthase inhibitor raltitrexed \cite{anderson2003process}, and the antibiotic norfloxacin \cite{rutenber1996binding}). Accelerating SBDD with deep learning can help address existing drug discovery challenges, such as polypharmacology by design \cite{reddy2013polypharmacology}, selectivity optimization \cite{kawasaki2011finding}, activity cliff prediction \cite{van2022exposing}, and target deorphanization \cite{civelli2013g}. Deep learning does not require explicit feature engineering and can thus be applied to learn directly from molecular representations of both ligands and proteins. This is particularly relevant for SBDD, where engineering numerical features for complex molecular entities like proteins \cite{kurgan2011structural} is inevitably more laborious than for small molecules\cite{consonni2009molecular}. Therefore, deep learning for SBDD bears an untapped potential to capture highly non-linear structure-activity relationships and has recently started to show its promise. Accurate protein structure prediction efforts like AlphaFold\cite{jumper2021highly,baek2021accurate} are expected to further accelerate computer-assisted SBDD. Deep learning for SBDD is still in its infancy but is moving forward at a fast pace, and its relevance in the years to come is expected to increase.

\begin{figure}[t]
 \centering
\includegraphics[width=15cm]{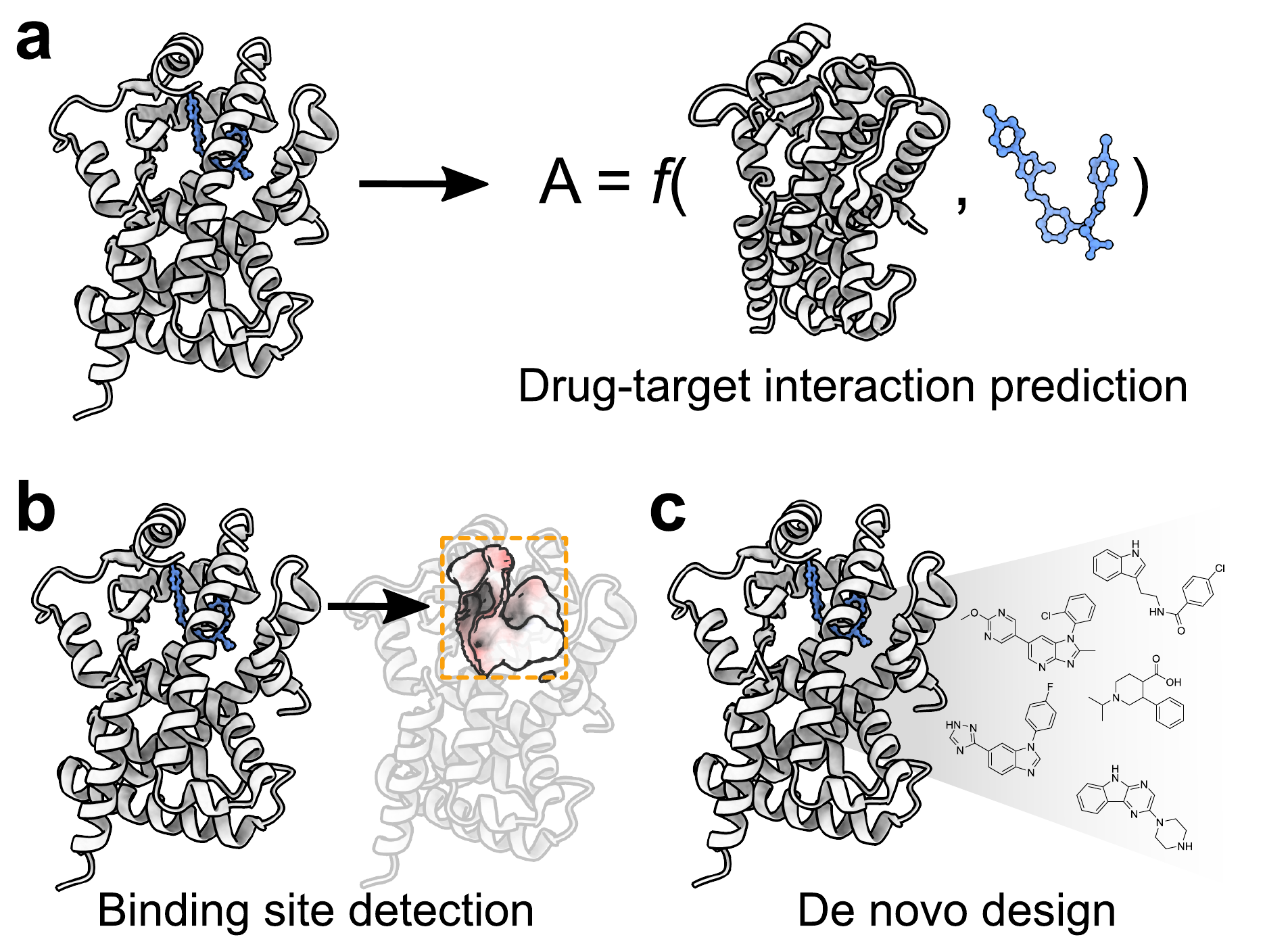} 
 \caption{\textbf{Structure-based drug discovery tasks discussed in this review}: (a) \textit{drug-target interaction prediction}, which aims to predict the affinity between a protein and a ligand, using the structural information of both molecular entities; (b) \textit{binding site detection}, which aims to identify `druggable' cavities in the protein structure, (c) \textit{de novo design}, aiming to design bioactive molecules from scratch using the information of a protein target.}
 \label{fig:summary}
 \end{figure}

This review provides a comprehensive overview of how deep learning can be leveraged for SBDD, to incorporate protein information at different levels of complexity (\textit{e.g.}, aminoacid sequence, and/or tertiary structure). After addressing how proteins can be represented for deep learning, we address current state-of-the-art methods for structure-based drug discovery, with a particular focus on drug-target interaction prediction, binding site detection, and \textit{de novo} design (Fig. \ref{fig:summary}). Finally, we discuss current limitations and research gaps, along with foreseen future directions and opportunities. A glossary of selected terms can be found in Box 1. 

\begin{tcolorbox}[float*=t, floatplacement=t, width=\textwidth]
\subsection*{Box 1: Glossary of selected terms}

\small{
\textbf{Binding site.} Protein region that is responsible for the interaction with another molecule (\textit{e.g.,} small molecule inhibitors, activators, and other proteins).\\
\textbf{Convolutional neural network (CNN).} Neural network architecture that learns from local information on grid-like data through convolution operations.\\
\textbf{Molecular docking.} Computational procedure used to predict the predominant three-dimensional binding mode(s) of a molecule w.r.t. another (macro)molecule it binds to. These typically involve the use of a conformational search method and pose scoring functions\cite{morris2008molecular}. \\
\textbf{Generative deep learning.} Deep learning methods that aim to model the underlying data distribution of a given set of samples and, by sampling from the modeled distribution, generate new data points without the need for explicit hard-coded design rules\cite{foster2019generative}.\\
\textbf{Geometric deep learning.} Umbrella term to identify neural network architectures that incorporate and process symmetry information in their design\cite{bronstein2021geometric}.\\
\textbf{Molecular descriptors.} Numerical features obtained from a molecular representation with the goal of capturing pre-defined chemical information\cite{consonni2009molecular}.\\
\textbf{Ligand.} Any molecule that binds to a protein with high affinity and specificity.\\
\textbf{Protein.} (Macro)molecule consisting of amino acid residues joined by peptide bonds.\\
\textbf{Reinforcement learning.} Subfield of machine learning whose goal is to study the agents that learn to act towards maximizing a cumulative reward in an environment.\\
\textbf{Simplified Molecular Input Line Entry System (SMILES)}\cite{weininger1988smiles}\textbf{.} String-based chemical notation capturing two-dimensional molecular information, in which letters are used to represent atoms, whereas symbols and numbers encode bond types, connectivity, branching, and stereochemistry.\\
\textbf{Transfer learning.} A machine learning method where a model trained on one task is reused as the starting point for a model on a second, related, task.\\
\textbf{Voxel.} A volumetric pixel.
}

\end{tcolorbox}

\section*{Representing proteins for deep learning}
\begin{figure}[t]
 \centering
\includegraphics[width=17cm]{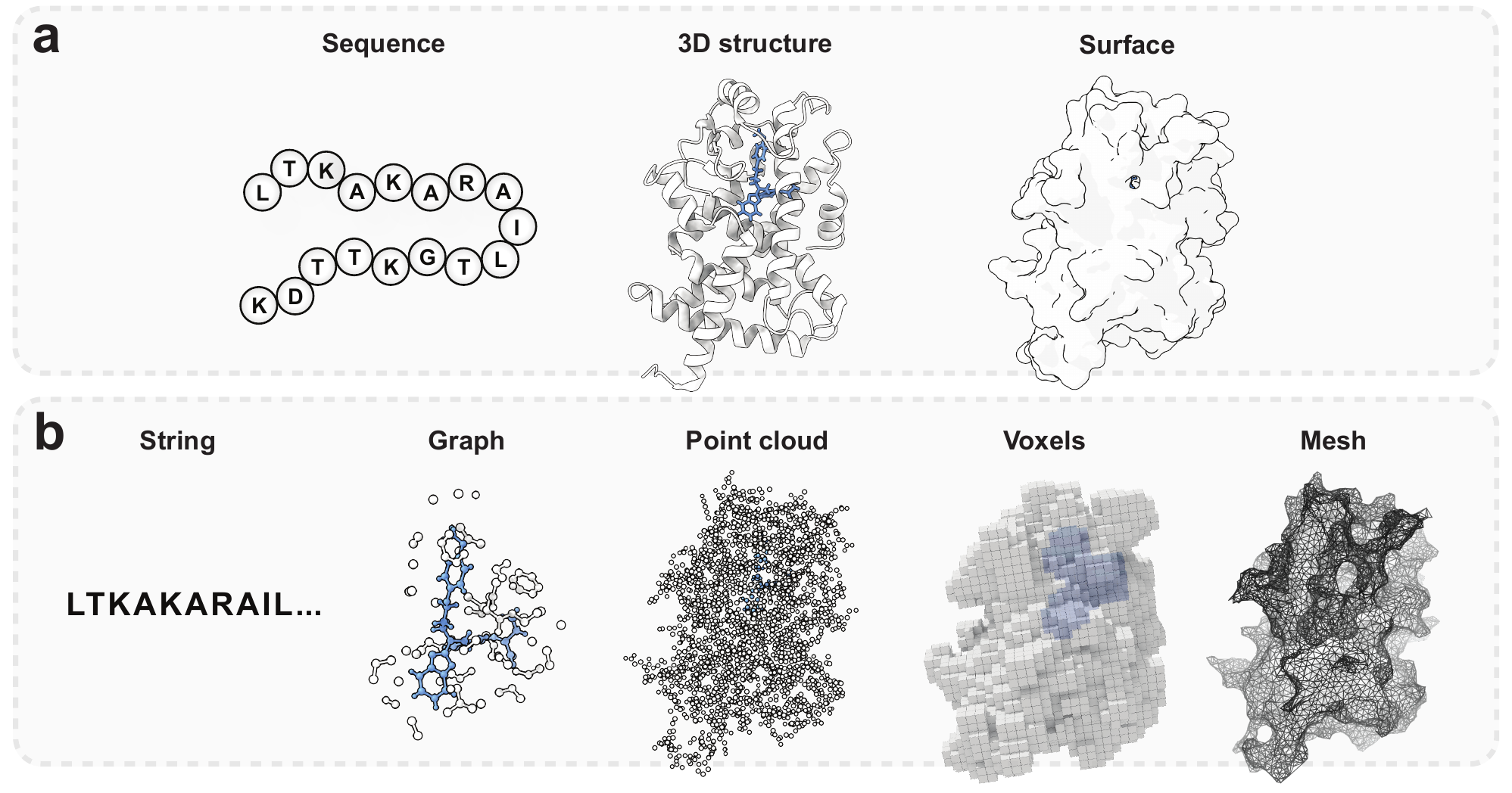} 
 \caption{\textbf{Representing proteins for deep learning. } (a) Structural hierarchy of protein information: \textit{(primary) amino-acid sequence}, referring to the sequential arrangement of amino acids along the polypeptide chain; \textit{(tertiary) 3D structure}, capturing protein folding in the three-dimensional space; \textit{protein surface}, delimiting solvent-accessible and inaccessible regions. Each level of information is characterized by a different availability of data (\ref{tab:dti_datasets}). (b) Protein representations for deep learning, capturing information on the protein sequence (\textit{strings}), the 3D structure (\textit{graphs}, \textit{voxels} and \textit{point clouds}), and the surface (\textit{point clouds} and\textit{ meshes}). Each representation is suited to different neural network architectures (Box 2).}
 \label{fig:molrepr}
 \end{figure}

The design of deep learning approaches for SBDD is inherently more intricate than those that are ligand-based, due to the need to represent protein information at different levels of complexity. Proteins are large polypeptide chains, organized in a `structural hierarchy'\cite{branden2012introduction}: (a) \textit{primary structure}, referring to the sequential arrangement of amino acids along the polypeptide chain, (b) \textit{secondary structure}, capturing the occurrence of alpha-helices and beta-pleated sheets along the protein sequence, and (c) \textit{tertiary structure}, capturing how proteins fold in the three-dimensional space. Such complexity is reflected in the various protein representations used for deep learning (Fig. \ref{fig:molrepr}): 
\begin{itemize}
\itemsep0em 
\item {\textit{Primary (amino acid) sequence}}. The amino acid sequence is specified starting from the amino-terminal end (N-terminus) and ending at the carboxyl-terminal (C-terminus) end. For deep learning purposes, the primary sequence is often represented as a character string, where each letter represents one of the twenty naturally-occurring amino acids (\textit{e.g.,} `AIR' corresponds to alanine, isoleucine, and arginine). These representations are at the core of established protein `featurization' techniques, such as ProtVec~\cite{asgari2015continuous}, SeqVec~\cite{heinzinger2019modeling}, and ProtTrans~\cite{elnaggar2020prottrans}. Although less frequently encountered~\cite{borgwardt2005protein, ingraham2019generative}, the primary sequence can also be represented as a graph, whose nodes are amino acids (featurized by type or corresponding physicochemical features) and whose edges capture their adjacency in the chain.

\item \textit{Tertiary (three-dimensional) structure}. The three-dimensional shape of a protein (tertiary structure) is determined by the interactions among its side chains, and features a certain degree of conformational plasticity~\cite{teague2003implications}. The protein structure contains key information for SBDD, since it relates to protein function \cite{orengo1999proteinfunction,orellana2019large}, and it determines ligand binding \cite{di2020mechanisms}. Moreover, inducing conformational changes is often the goal of drug discovery\cite{teague2003implications}. Several ways exist to learn from tertiary structures with deep learning. Early approaches~\cite{ragoza2017protein, jimenez2017deepsite} have used grid-based voxel representations (\textit{see} Box 1) to capture the spatial distribution of the protein's physicochemical or pharmacophore properties. While these representations are suited to well-established deep learning architectures (\textit{e.g.}, convolutional neural networks, CNNs), many voxels representing empty space do not carry relevant information, and they involve increased computational costs at higher spatial grid resolutions. Other approaches\cite{feinberg2018potentialnet} represent proteins as molecular graphs, where each atom is a node and each bond is an edge. Depending on the chosen level of coarse-graining, often only backbone atoms are chosen to correspond to nodes, while edges often represent geometrical proximity in the coordinate space rather than direct chemical bonds \cite{kong2022conditional}. Edges and bonds can be characterized by additional geometrical and/or physicochemical properties. 

\item {\textit{Protein surface}}. The protein surface is usually defined as the separation between solvent-accessible and inaccessible regions
\cite{lee1971interpretation,connolly1983analytical} (Fig. \ref{fig:molrepr}a) and it plays a key role in the protein interactions with (macro)molecular entities. Protein surfaces are usually represented as either meshes (\textit{i.e.}, a set of polygons capturing the location of the surface, whose vertices can be described by a 2D grid or a 3D graph structure) or point clouds (\textit{i.e.}, graphs whose nodes describe the location of the surface at a certain resolution). Although often computed from the 3D structure, the surface representation might better reflect the physicochemical features responsible for the interaction with other (macro)molecules, as well as aspects of protein function that go beyond sequence similarity.
\end{itemize}

The chosen protein representation for deep learning affects the \textit{type}, \textit{quality}, and \textit{quantity} of chemical information captured. It will also affect the type of suitable deep learning strategies  (Box 2) and the corresponding advantages and drawbacks. Additionally, the number of data available for machine learning depends on the chosen protein representation (Table \ref{tab:dti_datasets}). Primary sequences are easy to obtain but lack information about the spatial configuration of atoms, particularly relevant to determine the binding pose of ligands. On the other hand, 3D protein structures contain potentially richer information for drug discovery, but they are more difficult to obtain experimentally and thus relatively scarce (although the cost has been steadily decreasing with the advent of experimental techniques like cryogenic electron microscopy~\cite{renaud2018cryo}). In this context, AI breakthroughs in protein structure prediction like AlphaFold\cite{jumper2021highly} bear promise to bridge the gap in data availability, by making thousands of predicted protein structures available in an unprecedented effort\cite{varadi2022alphafold, hekkelman2021alphafill}. 
 
Small molecule ligands can be represented in analogous ways to protein structures. The most commonly used representations are: (a) molecular strings (\textit{e.g.,} SMILES strings\cite{weininger1988smiles}), which capture 2D information (atom occurrence and connectivity), (b) 2D and 3D molecular graphs (based on the availability of experimentally determined or computed conformational information), and (c) molecular surfaces. An in-depth description of small molecule representations and corresponding deep learning approaches can be found in a recent work~\cite{atz2021geometric}.

\begin{table}[ht!]\centering

\renewcommand{\arraystretch}{1.4}

\caption{\textbf{Summary of selected datasets for structure-based deep learning}. Dataset name, description and number of entries (updated as of June 2022) are provided.}
\label{tab:dti_datasets}
\begin{tabular*}{\textwidth}{llll} \toprule
\textbf{Dataset} &\textbf{Description} &\textbf{No. entries} & \textbf{Link (if available)}\\
\midrule

\makecell[tl]{\textit{Protein Data Bank}\\ (PDB)\cite{berman2003announcing}} & {\makecell[tl] {Structural data of biological\\ macromolecules.}} & 189,735 structures & \href{https://www.rcsb.org/}{rcsb.org}\\

{\textit{scPDB}}\cite{desaphy_2014_scpdb} & \makecell[tl]{Druggable binding sites and ligands  \\extracted from the PDB.} & \makecell[tl]{4782 protein structures\\ and 6326 ligands.} 

& \href{http://bioinfo-pharma.u-strasbg.fr/scPDB/}{bioinfo-pharma.u-strasbg.fr/scPDB}\\

{\textit{BioLip}}\cite{yang2012biolip} & \makecell[tl]{Semi-manually curated \\ ligand–protein interactions.} & 573,225 entries. 
& \href{https://zhanggroup.org/BioLiP/}{zhanggroup.org/BioLiP}\\

{\textit{PDBbind}\cite{wang2004pdbbind}} & \makecell[tl]{Protein-ligand co-complexes\\ and associated affinities extracted\\ from the PDB.} & 23,496 complexes. & \href{http://www.pdbbind.org.cn/}{pdbbind.org.cn} \\

{\textit{UniProt}\cite{wang2004pdbbind}} & \makecell[tl]{Protein sequence and functional\\ information.} & > 60 million sequences. & \href{https://www.uniprot.org/}{uniprot.org} \\

\makecell[tl]{{\textit{AlphaFold Protein}}\\ \textit{Structure Database}\cite{varadi2022alphafold}} & \makecell[tl]{Predictions of protein structures \\by AlphaFold v 2.0. \cite{jumper2021highly}} & \makecell[tl]{992,316 predicted\\ structures.} & \href{https://alphafold.ebi.ac.uk/}{alphafold.ebi.ac.uk}\\

\makecell[tl]{{\textit{AlphaFill}} \cite{hekkelman2021alphafill}} & \makecell[tl]{Common ligands and cofactor \\ transplants for AlphaFold models.} & \makecell[tl]{12,029,789 transplants.} & \href{https://alphafill.eu/}{alphafill.eu}\\

\makecell[tl]{\textit{Binding MOAD}\cite{benson2007moad,smith2019moad}} & \makecell[tl]{X-ray crystal structures with \\ bound ligands and experimental \\ binding affinities.} & \makecell[tl]{41,409 protein-ligand \\complexes, and
15,223 \\ binding measurements.} & \href{http://www.bindingmoad.org/}{bindingmoad.org}\\

\makecell[tl]{\textit{Directory of Useful}\\ {\textit{Decoys (DUD-E)}}\cite{mysinger2012dude}}
& {\makecell[tl] {Directory of decoys designed to\\ benchmark molecular docking\\ programs.}} & \makecell[tl]{22,886 active molecules\\and affinities on 102 \\ targets; 59 decoys per \\compound.} & \href{http://dude.docking.org/}{dude.docking.org}\\

\makecell[tl]{\textit{BindingDB}~\cite{liu2007bindingdb}\\ validation sets} & \makecell[tl]{Binding affinities of protein-ligand\\pairs curated from the literature.} & \makecell[tl]{$~\sim$1200 series with at least\\ 1 cocrystal available in each.} & \href{https://www.bindingdb.org/rwd/bind/index.jsp}{bindingdb.org} \\

\textit{BigBind}\cite{brocidiacono2022bigbind} & \makecell[tl]{Associated protein structures\\ to ChEMBL~\cite{gaulton2012chembl} assay data\\ via Pocketome.~\cite{kufareva2012pocketome}} & \makecell[tl]{818,995 activities with \\ associated protein structures.} & \href{https://chemrxiv.org/engage/chemrxiv/article-details/6384ad70c5675357f89943c5}{Brocidiacono \etal, 2022} \\

{\textit{KIBA}\cite{tang2014making}} & \makecell[tl]{Bioactivity measurements\\of compounds against kinases.} & 246,088 measurements. & \href{https://pubs.acs.org/doi/10.1021/ci400709d}{Tang \etal, 2014} \\

{\textit{Davis}\cite{davis2011comprehensive}} & \makecell[tl]{Binding affinities ($K_d$ values)\\ of inhibitors against kinases.} & 30,056 measurements.  & \href{https://www.nature.com/articles/nbt.1990}{Davis \etal, 2011} \\

\bottomrule
\end{tabular*}
\label{tab:datasets}
\end{table}

\section*{Deep learning for structure-based drug discovery}
This section aims to provide a concise overview of SBDD approaches fueled by deep learning. SBDD will be considered in its broader sense, \textit{i.e.}, not only limited to 3D protein structure, but also including sequence and surface representations. We focus on three key tasks (Fig. \ref{fig:summary}), namely binding site detection, drug-target interaction prediction, and structure-based \textit{de novo} design. For each task, selected deep learning approaches are described through the lenses of the protein representation they rely on (sequence, structure, or surface). A summary of selected deep learning studies is reported in Table \ref{tab:sbdd_summary2}.

\begin{table}[t]
\centering
\caption{\textbf{Selected deep learning approaches applied to SBDD.} Models are categorized by task and frequently adopted molecular representations.}

\label{tab:sbdd_summary2}
\begin{tabular}{lll}\toprule
\textbf{Task} &\textbf{Description} &\textbf{Protein representation } \\\midrule
{\makecell[tl] {\textit{Drug-target interaction}\\ \textit{prediction}}} & {\makecell[tl] {Predict the interaction between one or\\ more proteins and one or more ligands.}} & {\makecell[tl] {
Amino-acid sequence \cite{ozturk2018deepdta, karimi2019deepaffinity, shin2019self, gaspar2021proteochemometric, wen2017deep, cheng2022iifdti}\\
3D structure \cite{wallach2015atomnet, ragoza2017protein, zheng2019onionnet, lim2019predicting,cho2020interactionnet} 
}} \\
\textit{Docking} & \makecell[tl]{Determination of a ligand pose\\ within a target binding site.} & {\makecell[tl] {
3D structure \cite{stark2022equibind, zhang2022e3bind, corso2022diffdock} }}
\\
\textit{Binding site detection} &  {\makecell[tl]{Identification and/or localization of\\ functional protein binding sites.}} & {\makecell[tl] {
Amino-acid sequence \cite{tsubaki2019compound, karimi2019deepaffinity, gligorijevic2021structure, lee2022sequence}\\
3D structure \cite{jimenez2017deepsite, jiang2019novel, stepniewska2020improving, kandel2021puresnet, tubiana2022scannet} \\ 
Surface \cite{gainza2020deciphering, sverrisson2021fast, mylonas2021deepsurf}}} \\

\textit{De novo design} & \makecell[tl]{Generation of ligands with desired properties\\ conditioned on a protein.} & {\makecell[tl] {
Amino-acid sequence \cite{grechishnikova2021transformer, uludougan2022exploiting} \\
3D structure \cite{skalic2019ligvoxel, ragoza2022generating, li2021structure, mcnaughton2022novo, chan20223d, skalic2019target, zhu2022pgmg, krishnan2021novo, zhang2022novo, schneuing2022structure, igashov2022equivariant} }}
\\
\bottomrule
\end{tabular}
\end{table}

\subsection*{Drug-target interaction prediction}

The identification of interactions between molecules and macromolecular targets is a key step in drug discovery, drug repurposing, and off-target activity prediction. Drug-target interaction (DTI) prediction aims to predict the bioactivity (\textit{e.g.}, binding affinity) of a given set of molecules on one or more macromolecular targets, by leveraging both protein and ligand information. Given the complexity of ligand-protein interactions and of engineering suitable molecular features for DTI, it is no surprise that this topic has found a widespread application of deep learning techniques \cite{bagherian2020dti}. In what follows, deep learning models developed for DTI prediction are categorized on the basis of the protein representation they rely on.

\begin{itemize}
   \item{\textit{Sequence-based approaches}}.
    Sequence-based DTI prediction models use amino acid sequences in combination with ligand representations to perform a prediction. One of the earliest approaches, DeepDTA\cite{ozturk2018deepdta}, applied 1-dimensional CNNs to simultaneously learn from string representations of both ligands  (in the form of SMILES) and protein sequence, by first creating separated embeddings and then concatenating them to perform a prediction. Later works have replaced CNNs with other methods, such as recurrent neural networks (Box 2),\cite{karimi2019deepaffinity,abbasi2020deepcda} and attention-based\cite{karimi2019deepaffinity, abbasi2020deepcda, zhao2022attentiondta} or transformer architectures\cite{shin2019self, monteiro2022dtitr, gaspar2021proteochemometric}. Several works have addressed how to improve the representation of protein sequences, \textit{e.g.}, by incorporating evolutionary information,\cite{ozturk2019widedta,ozccelik2021chemboost} or protein sequence composition descriptors \cite{wen2017deep,feng2018padme}. Ligands are usually represented as strings (\textit{e.g.}, SMILES \cite{ozturk2018deepdta, ozccelik2021chemboost} or DeepSMILES \cite{o2018deepsmiles,ozturk2019widedta}), binary fingerprints,\cite{wen2017deep,lee2019deepconv} frequently occurring substructures\cite{ozturk2019widedta}, or molecular graphs.\cite{feng2018padme, nguyen2021graphdta, wang2022yuel}

    \item {\textit{3D structure-based approaches}}. These models leverage atom coordinates, usually of co-crystallized protein-ligand complexes, for training. Early approaches projected 3D protein-ligand complexes into grids featurized with physio-chemical properties, and subsequently applied CNNs for binding affinity prediction\cite{wallach2015atomnet, ragoza2017protein, jimenez2018k}. Later works extended this idea by including more sophisticated features, \textit{e.g.}, inter-molecular interaction fingerprints\cite{zheng2019onionnet} and computed molecular energies\cite{hassan2020rosenet}. 3D grid-based approaches have also been used for lead optimization by predicting relative binding free energies linked to small modifications of ligand structures\cite{jimenez2019deltadelta, mcnutt2022improving}. More recent approaches have replaced grid-based representations with graphs \cite{feinberg2018potentialnet, torng2019graph, lim2019predicting, cho2020interactionnet}, allowing to explicitly represent atom neighborhoods and connectivity, and apply roto-translational invariant graph neural networks for binding affinity prediction. 
    
    \item \textit{Surface-based approaches}. Surface-based approaches have found limited application for DTI prediction. OctSurf\cite{liu2021octsurf} represents both binding pockets and ligands as surfaces, by partitioning the 3D space recursively into octants and considering only portions containing van der Waals surface points. Non-empty octants, along with their physicochemical and geometric features, are then used as the input to a CNN. Other approaches, such as HoloProt\cite{somnath2021multi}, merge 3D structure (graph) and surface (point cloud) information for task-specific training, \textit{e.g.,}  enzyme-catalyzed reaction classification and binding affinity prediction.

\end{itemize}

Another topic of recent interest by the deep learning community is \textit{protein-ligand docking}, which aims to predict the putative binding pose of a ligand upon binding to a macromolecular target (Box 1). Although these methods do not aim to predict the affinity between a ligand-protein pair directly, they can be used as a proxy to elucidate potential mechanisms of interaction. Deep learning has been mostly applied to ligand pose optimization while considering a rigid target structure, although recent approaches have started taking side-chain flexibility into account~\cite{qiao2022dynamic}. 
Early approaches used protein-ligand interaction fingerprints\cite{pereira2016boosting, gentile2020deep}, while successive approaches have leveraged either a voxelized version of the protein structure combined with CNNs~\cite{wang2020protein, mcnutt2021gnina} or graph-based representations with message-passing neural networks\cite{zhang2019deepbindrg,morrone2020combining,mendez2021geometric,stafford2022atomnet} in lieu of classical scoring functions. 
Finally, several approaches have attempted to directly predict the ligand binding pose in an end-to-end fashion~\cite{masters2022deep,stark2022equibind, zhang2022e3bind}, without the need for a classical search algorithm by exploiting advances in equivariant deep learning.

Deep learning has undoubtedly accelerated DTI prediction, thanks to the possibility to represent and learn from protein-ligand complexes more efficiently. However, %
simpler models based on well-established descriptors might reach comparable performance \cite{volkov2022frustration}, due to undesired memorization and hidden bias in ligand-protein interaction data\cite{peng2003exploring,chaput2016benchmark,chen2019hidden,tran2020lit}. Moreover, no relationship has been observed between the complexity of protein and ligand representations and the accuracy of the resulting deep learning models \cite{volkov2022frustration}. Thus, more attention should be put on strategies for model evaluation and data selection/splitting procedures to ensure a reliable prediction of DTIs with deep learning \cite{volkov2022frustration,ozccelik2022debiaseddta,zhu2022sbvs}.

\subsection*{Binding site detection}

The identification of `druggable' binding sites in proteins plays a pivotal role in SBDD, from hit identification and molecule screening to mechanism formulation\cite{perot2010druggable}. Over the years, a plethora of methods have been developed for binding site detection\cite{perot2010druggable,zhao2020lbsreview, laskowski1995surfnet, brady2000fast, capra2009predicting}, \textit{e.g.}, via interatomic gap volumes\cite{laskowski1995surfnet} or regions of buried pocket surfaces\cite{weisel2007pocketpicker}. Recently, deep learning methods learning directly from `raw' representations of proteins have gained increasing traction to detect binding sites. These approaches can be grouped by the molecular representations they rely on, \textit{i.e.}, protein sequence, 3D structure, and surface, as described below:

\begin{itemize}
    \item \textit{Sequence-based models}. Binding site detection can be performed by predicting which residues of the amino-acid sequence are involved in ligand binding, although sequence-based approaches have found limited application. Early methods approached binding site detection as a `side-product' of binding affinity prediction, by using explainable AI techniques to highlight relevant residues for a model's prediction \cite{tsubaki2019compound,karimi2019deepaffinity,gligorijevic2021structure}. Few works have addressed binding site detection only \cite{cui2019predicting,khan2022prob}. Recently, sequence-based binding site detection has been jointly modeled with drug-target affinity prediction, leading to improved performance on both tasks~\cite{lee2022sequence}. 
    
    \item \textit{3D structure-based models}, which use the spatial information of proteins to detect likely binding sites. Early approaches represented the protein structure with voxels featurized with pharmacophore-like properties, along with convolutional neural networks\cite{jimenez2017deepsite,jiang2019novel}. Subsequent works have refined structure-based binding site detection with additional techniques from the computer vision domain, \textit{e.g.}, image segmentation.\cite{stepniewska2020improving, kandel2021puresnet}. BiteNet\cite{kozlovskii2020spatiotemporal} extended `static' CNN-based approaches by incorporating conformational ensembles of proteins. The approach was later adapted to predict protein-peptide binding sites \cite{kozlovskii2021protein}. A recent approach combined spatial properties with amino acid sequence information to predict protein-protein interaction sites and is applicable to ligand binding site prediction\cite{tubiana2022scannet}.
   
    \item \textit{Surface-based models}. Voxelized representations of protein coordinates have several drawbacks,\cite{mylonas2021deepsurf,krivak2018p2rank} (\textit{e.g.}, carrying non-informative voxels that are either deeply buried and not accessible by a ligand or represent empty space, and information coarse-graining due to discretization of the input protein space) and might lead to worse results than working with surfaces alone\cite{krivak2018p2rank}. For this reason, several methods based on protein surfaces have been developed over the years. These approaches rely on the representation of protein structures as continuous shapes characterized by geometric and physicochemical features to perform a prediction. Geodesic CNNs have been used to determine interaction fingerprints of molecular surfaces, and to predict protein and ligand binding sites \cite{gainza2020deciphering}. The approach was later expanded to obtain fully-learnable protein representations.\cite{sverrisson2021fast} DeepSurf\cite{mylonas2021deepsurf} discretizes the solvent accessible surface using a combination of \textit{K}-means clustering and and density reduction\cite{mylonas2021deepsurf}. 
    
\end{itemize}

A recent analysis of computational approaches for protein-ligand binding site recognition~\cite{gagliardi202220} has shown DeepSurf\cite{mylonas2021deepsurf} to perform remarkably well. Moreover, non-machine-learning algorithms resulted competitive alternatives to deep learning \cite{gagliardi202220}. All analyzed methods struggled on shallow binding sites, due to the higher frequency of deep grooves used for model training \cite{gagliardi202220}. Despite the recent progress of binding pocket detection, room for improvement remains, \textit{e.g.}, to increase pocket coverage and detect sub-pockets\cite{gagliardi202220}, and to predict allosteric binding sites \cite{ni2022allostery}.

\subsection*{Protein-based \textit{de novo} design}

\textit{De novo} design refers to the generation of novel chemical entities possessing desired properties from scratch \cite{schneider2016denovo} and is among the most challenging tasks in computer-assisted drug discovery. Computational algorithms are faced with an incredibly vast `chemical universe', whose cardinality has been estimated between 10$^{24}$ and 10$^{100}$ molecules ~\cite{ertl2003cheminformatics, dobson2004chemical,lipinski2004navigating}. In this context, `brute-force' molecule assembly or enumeration approaches are computationally unfeasible. In recent years, generative deep learning has shown great promise for \textit{de novo} drug design \cite{segler2018generatingfocused,merk2018denovo,olivecrona2017molecular} and to complement traditional approaches based on human-engineered rules \cite{devi2015evolutionary, douguet2000genetic,anderson2003process, ferreira2015molecular}.

Generative deep learning approaches for \textit{de novo} design are usually applied to produce molecules in the form of molecular graphs\cite{li2018learning,samanta2020nevae,de2018molgan} or strings\cite{segler2018generatingfocused, yuan2018chemicalspacemimicry, merk2018denovo} (\textit{e.g.}, SMILES). While most \textit{de novo} design approaches are ligand-based \cite{olivecrona2017molecular,popova2018deep,staahl2019deep,segler2018generatingfocused,gupta2018generative,moret2020generative}, structure-based approaches have recently emerged as a promising research direction, due to their potential to design molecules interacting with pharmacologically-relevant targets on demand. 

\begin{itemize}
     \item \textit{Sequence-based approaches.} Sequence-based \textit{de novo} design approaches usually cast the problem into a machine translation task, where high-affinity protein-ligand pairs are considered as sentences in different languages to be matched. To this end, sequence and SMILES strings for proteins and ligands are used, respectively. The first-in-kind approach \cite{grechishnikova2021transformer} trained a transformer architecture to `translate' amino-acid sequences into the SMILES strings of the corresponding ligand. This approach can be used for sequence-conditioned \textit{de novo} design. A recent work used a transformer-based pipeline \cite{uludougan2022exploiting} that combined language models that were pre-trained on large corpora of proteins \cite{filipavicius2020pre} and small molecules \cite{chithrananda2020chemberta}. 
     
    \item \textit{3D structure-based approaches}. \textit{De novo} design conditioned on the tertiary structure information can usually generate molecules in the form of 3D ligands (molecular graphs) or strings (\textit{e.g.}, SMILES). In the former case, 3D representations of protein-ligand complexes are used as the input to generate novel 3D molecular graphs. As one of the earliest approaches, LigVoxel\cite{skalic2019ligvoxel} relied on 3D grids to generate spatial `blobs' of ligand properties such as occupancy, aromaticity, and hydrogen-bond donor/acceptors that match the protein pocket. Later works used equivariant diffusion networks\cite{schneuing2022structure, igashov2022equivariant}, variational autoencoders\cite{ragoza2022generating}, and reinforcement learning\cite{mcnaughton2022novo,li2021structure} to directly generate ligand conformations for the binding pocket. Recently, equivariant neural networks coupled with point-cloud representations have been used for molecule optimization, via pocket-based fragment expansion \cite{powers2022fragment}.  Compared to 3D graphs, molecular strings are usually easier to generate and might match or outperform graph-based models~\cite{flam2022language}. A pioneering work of this category leverages generative adversarial networks\cite{goodfellow2014generative} to produce SMILES strings for protein pockets, where a pocket is featurized and fed into a generator network with a conditioning ligand\cite{skalic2019target}. A subsequent model adopts graph neural networks to represent active sites and generates targeted SMILES strings\cite{krishnan2021novo}, whereas another model uses a pharmacophore model to represent ligands and condition the generation for the targeted pocket\cite{zhu2022pgmg}. Recently~\cite{zhang2022novo}, a recurrent neural network model has been coupled with ligand–protein interaction fingerprints (determined on ligand docking poses) for conditioned ligand generation in the form of SMILES strings.
    
\end{itemize}

While ligand-based \textit{de novo} design pipelines using deep learning have been experimentally validated in multiple instances \cite{merk2018denovo,grisoni2021combining,moret2021beam}, structure-based \textit{de novo} design has to date not been applied prospectively. This highlights an important gap in the potential for acceptance of structure-based \textit{de novo} design algorithms. 

\begin{tcolorbox}[floatplacement=t, width=\textwidth]
\subsection*{Box 2: Simplified depiction of common neural networks applied to protein representations.\\}

Each protein representation useful for SBDD (Fig. \ref{fig:molrepr}) is suited to different neural network architectures. Well-established examples of such architectures are the following:

\begin{itemize}
\item[a.] \textit{Recurrent neural networks} (RNNs) (\textit{e.g.}, with long-short term memory cells \cite{schmidhuber1997long}) are commonly-chosen architecture to process the primary sequence of a protein. RNNs incorporate feedback connections that allow information in the previous inputs to flow into the subsequent inputs. The feedback mechanism behaves as a `learned memory' in the architecture and enables capturing long-range dependencies in the input sequence. %
\item[b.] \textit{Convolutional neural networks} (CNNs) are a powerful architecture when paired with voxelized representations to capture spatial dependencies. CNNs apply learnable filters to the input and excel in capturing local patterns, which renders them suited to binding affinity prediction and binding site detection.

\item[c.] \textit{Graph neural networks} (GNNs) operate on molecular graphs (\textit{e.g.}, atoms and their interactions) and can capture the structural and functional relationships between, as well as within, atoms belonging to one or more molecular complexes. The wide application domain of graphs enables GNNs to find applications in every task of structure-based drug discovery. %
\end{itemize}

\centering
\includegraphics[width=15cm]{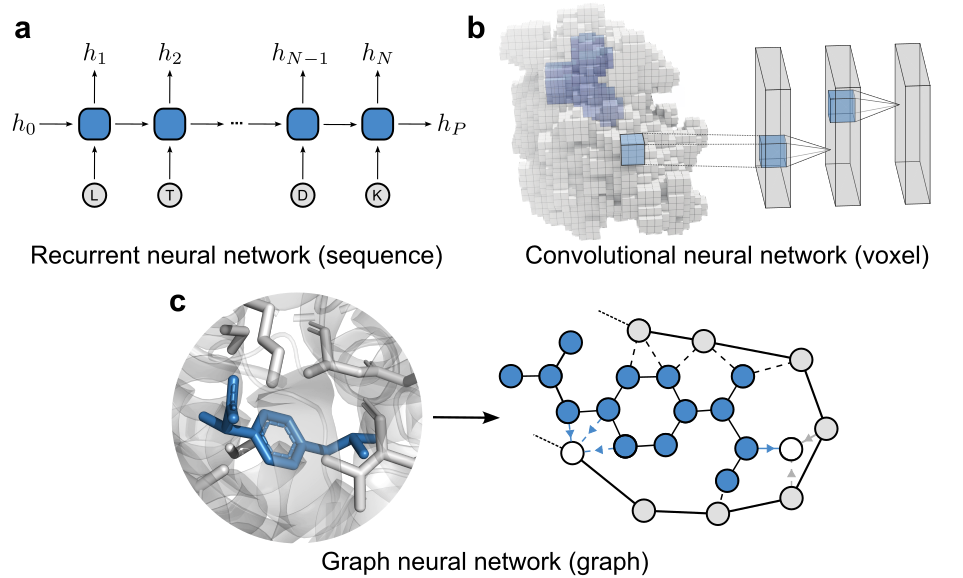}

\vspace{1em}

\end{tcolorbox}

\section*{Gaps, opportunities, and outlook}
Deep learning for structure-based drug discovery is gaining increasing traction, as evidenced by the exponential increase in the number of published approaches over the last few years. Breakthroughs in protein structure prediction\cite{varadi2022alphafold,baek2021accurate} not only exemplify the potential of deep learning in the molecular sciences but are expected to further propel structure-based drug discovery with AI. SBDD is arguably more challenging than its ligand-based counterpart, due to the number and structural complexity of chemical entities involved, and aspects like target conformational flexibility~\cite{hatzakis2014single}. Most of the current SBDD algorithms are agnostic to dynamical information and, in certain instances, might not outperform simpler methods~\cite{gagliardi202220,volkov2022frustration}. However, SBDD has an untapped potential to tackle drug discovery for new, uncharted protein targets. Such a `zero-' and `few-shot' learning potential makes AI-driven SBDD a high-risk/high-gain endeavor, expected to advance future drug discovery.

An area of continuous development that bears promise for SBDD is geometric deep learning\cite{bronstein2017geometric, bronstein2021geometric}. Geometric deep learning attempts to unify neural networks from the perspective of symmetry and topology. Roto-translational invariance/equivariance is particularly relevant for the three-dimensional representations of molecular systems, which is beneficial to limit function search space during training~\cite{atz2021geometric}. Geometric deep learning is expected to boost AI's capability to model molecular complexes and their interactions\cite{gainza2020deciphering,zhang2022efficient}, as well as molecular design\cite{igashov2022equivariant,gainza2022novo}. Diffusion models \cite{sohl2015deep} -- a family of generative models inspired by non-equilibrium thermodynamics -- are also gaining increasing popularity in deep learning thanks to their generative capabilities, and have found pioneering applications in the molecular sciences, too \cite{igashov2022equivariant,hoogeboom2022equivariant,corso2022diffdock}. These approaches have reached state-of-the-art in several deep learning applications and are expected to propel SBDD in the future.

A current bottleneck of AI-driven SBDD is constituted by the available training data. While protein sequence information and experimental assay data are largely available, high-quality 3D data of co-crystallized proteins and ligands with accompanying properties is largely missing. Furthermore, obtaining high-quality protein structure information is resource-intensive and challenging for several drug targets\cite{davis2003crystallograpy}, \textit{e.g.}, disordered and membrane proteins. Furthermore, non- and poorly binding molecules are often strongly under-reported in medicinal chemistry datasets. As a result, available three-dimensional datasets are often highly biased in their content~\cite{peng2003exploring, chaput2016benchmark, chen2019hidden, tran2020lit, shah2020pitfalls}, which has historically led to poor generalizability~\cite{sieg2019need,boyles2020learning,wallach2018most,gonczarek2018interaction}. Several studies have attempted to alleviate this issue, via data curation\cite{wallach2018most, scantlebury2020data}, bias-controlled training\cite{chan20223d}, and debiasing\cite{ozccelik2022debiaseddta, sundar2019effect}.  Bridging the gap between the different types of available information will be an active task in upcoming years, with some recent work pointing in this direction already~\cite{brocidiacono2022bigbind}.

Finally, the application of deep learning in well-established fields like chemical biology and medicinal chemistry might at times be met with skepticism by experimentalists. These well-grounded concerns commonly originate from the black-box nature of deep learning. Additionally, robust performance benchmarks and evaluation datasets are currently missing, especially for \textit{de novo} design studies. Moreover, to the best of our knowledge, no deep learning approaches for SBDD have been validated in a prospective setting yet. To foster broader acceptance of structure-based deep learning, we need to `open the box' and validate methods experimentally. We envision that more sophisticated applications of explainable AI\cite{goebel2018xai42,jimenez2020drug} will aid in identifying underlying structure-activity relationships and binding modes, and bridging the gap between theory and real-world applications.

\section*{Conclusions}
In recent years, deep learning has taken drug discovery by storm, offering new opportunities for more efficient exploration of chemical space. Ligand binding site detection, drug-target interaction prediction, and structure-based de novo design can be valuable tools in early drug discovery, especially for unexplored macromolecular targets. As a whole, these approaches bear great promise to extend upon the successes of ligand-based methods. However, structure-based methods have not yet proven their applicability in prospective scenarios. Overcoming such barriers will mostly depend on additional efforts in data collection and curation, as well as on methodologies that efficiently exploit relationships between assay and structural data.

\section*{Acknowledgements}
F.G. acknowledges the support from the Institute for Complex Molecular Systems (ICMS, TU/e) and the Centre for Living Technologies (Alliance TU/e, WUR, UU, UMC Utrecht). 
\section*{Conflict of interest}
None to declare. 
\section*{Author contribution}\textit{Conceptualization}: F.G., with contributions from all authors; \textit{Investigation}: all authors; \textit{Visualization}: D.v.T., R.O., F.G.; \textit{Writing – original draft}: F.G., R.O., D.v.T.; \textit{Writing – review \& editing}: all authors.

\section*{List of abbreviations}

\textbf{AI}: Artificial intelligence.\\
\textbf{CNN}: Convolutional neural network.\\
\textbf{DTI:} Drug-target interaction prediction.\\
\textbf{GNN}: Graph Neural Network.\\
\textbf{SBDD}: Structure-based drug discovery.\\
\textbf{SMILES}: Simplified molecular input line entry system.\\
\textbf{RNN}: Recurrent Neural Network.\\
\textbf{2D (3D)}: Two- (three-)dimensional.\\

\newpage
\bibliography{ref.bib}

\end{document}